\documentclass[english,journal=jacsat,manuscript=article,layout=twocolumn]{achemso}
\usepackage[T1]{fontenc}
\usepackage[latin9]{inputenc}
\usepackage{color}
\usepackage{textcomp}
\usepackage{amstext}
\usepackage{graphicx}

\makeatletter

\usepackage{textcomp}

\usepackage{amstext}

\usepackage{esint}

\makeatletter

\author{Francesco Colizzi}
\email{colizzi@sissa.it}
\affiliation[SISSA]
{SISSA - Scuola Internazionale Superiore di Studi Avanzati, \\ via Bonomea 265, 34136 Trieste, Italy}
\author{Giovanni Bussi}
\email{bussi@sissa.it}
\affiliation[SISSA]
{SISSA - Scuola Internazionale Superiore di Studi Avanzati, \\ via Bonomea 265, 34136 Trieste, Italy}

\title[\texttt{achemso} Step-by-step RNA unwinding]
{RNA unwinding from reweighted pulling simulations}

\makeatother

\makeatother

\usepackage{babel}

\begin{document}
\begin{abstract}
The forming and melting of complementary base pairs in RNA duplexes
are conformational transitions required to accomplish a plethora of
biological functions. Yet the dynamic steps of these transitions have
not been quantitatively characterized at the molecular level. In this
work, the base \textcolor{black}{opening} process was first enforced
by atomistic pulling simulations and then analyzed with a novel reweighting
scheme which allowed the free-energy profile along any suitable reaction
coordinate, e.g.~solvation, to be reconstructed. The systematic application
of such approach to different base-pair combinations provides a molecular
motion picture of helix opening which is validated by comparison with
an extensive set of experimental observations and links them to the
enzyme-dependent unwinding mechanism. The RNA intrinsic dynamics disclosed
in this work could rationalize the directionality observed in RNA-processing
molecular machineries. 
\end{abstract}
\maketitle

\section{Introduction}

The ability of ribonucleic acid (RNA) to adopt peculiar three-dimensional
structures that mediate a variety of biological functions makes it
the most versatile regulatory factor in the cell.\cite{bloo+00book}
Virtually involved in all cellular processing of the genetic information,
the RNA is able to achieve such a functional diversity by adaptively
acquiring very distinct conformations in response to specific conditions
of the cellular environment.\cite{hall+11acr} Among the structural
rearrangements engaged by RNA, the opening of complementary base pairs
is an ubiquitous process required to accomplish a wide range of metabolic
activities such as transcription, pre-mRNA splicing, ribosome biogenesis
or translation initiation.\cite{roca-lind04nrmcb} In the cell this
is usually catalyzed by enzymes called RNA helicases which have been
shaped by the evolution to unwind double-stranded (ds) RNA according
to its intrinsic dynamic properties.\cite{jank-fair07cosb,pyle08arb}

From a molecular standpoint, the opening and forming of individual
base pairs are fundamental, yet poorly understood, events which provide
the structural framework to large-scale RNA conformational transitions
and folding.\cite{norb-nils02acr,zhua+07nar,li+08arb,alha-walt08cosb,oroz+08cosb,xia08cocb,hall+11acr,rinn+11acr}
In this respect and related to the work presented herein, insightful
investigations have
been reported only for short
deoxyribonucleic acids (DNAs)
in the B-form helical geometry.\cite{varn-lave2002jacs,haga+03pnas,giud-lave2003jacs}
Using transition-path sampling, Hagan et al\cite{haga+03pnas} have
fully characterized the energetics of (un)pairing for a 5\textasciiacute{}-end
cytosine. However, the mechanism underlying the complete opening of
the duplex has not been systematically faced nor analyzed. Moreover,
differences in topology and thermodynamic parameters between B-form
DNA and A-form RNA suggest that the mechanism of duplex \textcolor{black}{separation}
might obey different rules.

Recently, combining thermodynamic information with the relative population
of unpaired \textcolor{black}{terminal} nucleotides (dangling ends)
observed in large ribosomal RNA (rRNA) crystal structures, Mohan et
al\cite{moha+2009jpcb} have proposed that stacking and pairing reactions
are not simultaneous, and that 3\textasciiacute{}-single-strand
stack leads the base pairing of the 5\textasciiacute{}-strand.
Nevertheless, collecting an unbiased data set of dangling-end population is not trivial
and, when viewed in the context of the full ribosomal assemblies,
the single stranded (ss) regions are seen to interact extensively
with other RNA elements.\cite{moha+2009jpcb}
On top of that,
since the
formation and opening of base pairs is a dynamic process, both the
ensemble-averaged thermodynamic properties and the detailed but static
X-ray picture have to be complemented with other methods able to directly
and quantitatively capture the dynamics of the investigated event.
Likely, this gap will be efficiently bridged by \emph{ad hoc} designed
spectroscopic approaches.\cite{alha-walt08cosb,lee+10pnas,rinn+11acr,hall+11acr}
For instance, femtosecond time-resolved fluorescence spectroscopy
is emerging as a powerful technique for the quantitative analysis
of base-stacking pattern and base motion,\cite{zhao-xia09methods}
although its applications to probe RNA dynamics are still in their
infancy and the method presents several limitations.\cite{xia08cocb}
As a matter of fact, the integration of spectroscopic approaches with
other powerful techniques is presently needed to gain molecular details
on the RNA intrinsic dynamics.
\textcolor{black}{
Among the possible methodological choices,
atomistic simulations\cite{karp-mcca2002natsb}
allow any base sequence to be characterized
and all the microscopic parameters to be controlled.
Additionally, when
combined with state-of-the-art free-energy methods,
they can provide an unparalleled perspective on the
mechanism and dynamics of the biomolecular process of interest. As
far as the reconstruction of free-energy profiles is concerned, the
capability of estimating those profiles along any suitable reaction
coordinate, without any further computational cost, would offer researchers
a powerful and versatile tool enabling both the disclosure of intermediate
states and the multifaceted analysis of complex conformational transitions. }

With this spirit, here we report an \emph{in silico} study elucidating
the mechanism for
\textcolor{black}{strand separation} in the RNA double
helix. In particular, we used atomistic steered molecular dynamics
(MD) simulations\cite{soto-schu07science} to enforce the \textcolor{black}{unbinding}
of nucleobases into the surrounding explicit water. To allow a systematic
analysis of different base sequences we devised a novel Jarzynski-equation-based
reweighting scheme which allowed the free-energy landscape to be reconstructed
as a function of different reaction coordinates and the unbinding
energies to be straightforwardly estimated. The computed free-energy
differences are consistent with experimental observations and suggest
that the \textcolor{black}{strand separation} mechanism occurs by a
stepwise process in which the probability of unbinding of the base
at the 5\textasciiacute{} \textcolor{black}{terminus} is significantly
higher than that at the 3\textasciiacute{} \textcolor{black}{terminus.}
The biological implications of these findings are discussed \textcolor{black}{and
related to the unwinding mechanism catalyzed by RNA processing machineries.}
\textcolor{black}{Given the general nature of our approach,} the introduced
methodology can be directly applied to analyze a broad range of molecular
unbinding processes.

\section{Methods}

\begin{figure}
\includegraphics[width=1\columnwidth]{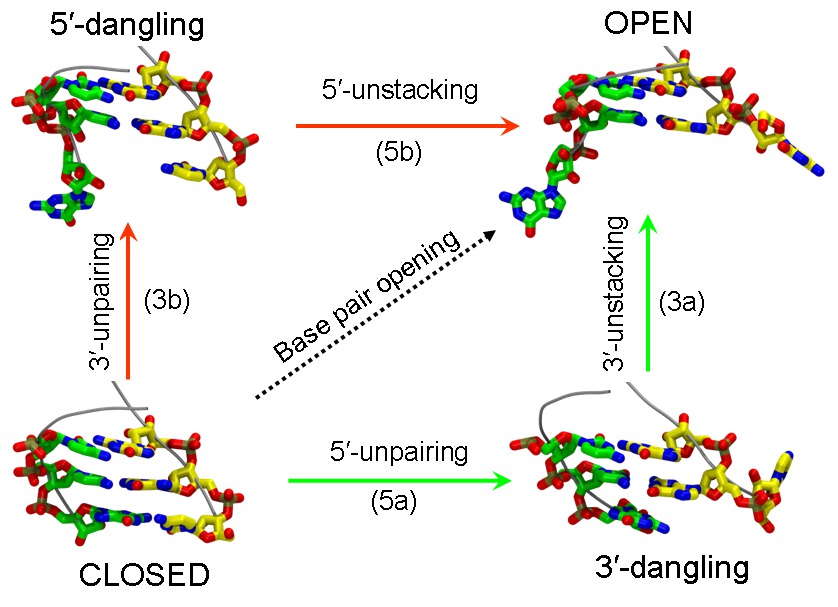}

\caption{\label{fig:Thermodynamics-cycle}\textcolor{black}{Elementary steps
involved in the opening of a base pair. The thermodynamics cycle was
used to characterize different base pair combinations.}}
\end{figure}

\textcolor{black}{Throughout the manuscript the following nomenclature
will be consistently used to define each elementary step involving
one single base and occurring during the opening of a closed base pair:
unpairing is used to define the process undergone by a single base
for which both Watson-Crick hydrogen bonds and stacking interactions
with adjacent bases are broken; unstacking is the process breaking
the stacking interactions between a dangling terminal nucleotide and
its adjacent bases. The opening of a base pair is thus composed by
an initial unpairing followed by an unstacking (}\ref{fig:Thermodynamics-cycle}\textcolor{black}{).
In the manuscript we also use the term unbinding referring to both
unpairing and unstacking processes by no means of specificity.
Finally, the strand with the 5\textasciiacute{} terminal (or 3\textasciiacute{} terminal) nucleobase
being pulled
is referred to as the  5\textasciiacute-strand (or 3\textasciiacute-strand).
}

\subsection{System set-up}

We simulated the unpairing and unstacking of nucleobases at both 3\textasciiacute{}-
and 5\textasciiacute{}-\textcolor{black}{termini} in dsRNAs of sequence
${\textrm{5\textasciiacute-CCGGGC-3\textasciiacute}\atop \textrm{3\textasciiacute-GGCCCG-5\textasciiacute}}$
and ${\textrm{5\textasciiacute-GGCCCG-3\textasciiacute}\atop \textrm{3\textasciiacute-CCGGGC-5\textasciiacute}}$
(\ref{fig:double-helix}). \textcolor{black}{Two sets of data can be
obtained from each dsRNA thus resulting }in four systems with different
combinations of Watson-Crick base pairing and stacking (\ref{fig:double-helix}A).
Both terminal and non-terminal base pairs (i.e.~a base pair at the
ss-ds RNA junction) were investigated. Non-terminal base pairs showed
the same trend in relative stability observed for terminal ones, and
are reported in the Supporting Information (SI). The A-form dsRNA
was built using ASSEMBLE\cite{jossi+10bionf} and then solvated with
$\sim$3600 water molecules, 20 Na$^{+}$ and 10 Cl$^{-}$ ions, resulting
in an excess salt concentration of about 0.15 M. The mobility of added
ions was fairly diffusive during the simulations. After minimization
and thermalization, each \textcolor{black}{system} \textcolor{black}{(or
intermediate)} was then evolved for 30 ns in the isothermal-isobaric
ensemble (300K, 1Atm)\cite{buss+07jcp,parr-rham81jap} using the Amber99
force field\cite{wang+00jcc} and TIP3P water.\cite{jorg+83jcp} Preliminary
calculations carried out using the recent refinement of the Amber99
force field (parmbsc0)\cite{pere+07bj} have shown quantitatively
similar results in the reconstructed free-energy profiles. This is
probably due to the poor involvement of the refined $\alpha$ and
$\gamma$ dihedrals during the unbinding trajectories. Long-range
electrostatic interactions were calculated with the particle mesh
Ewald method.\cite{dard+93jcp} Plain MD and biased steered MD trajectories
were generated with GROMACS 4.0.7\cite{hess+08jctc} combined with
PLUMED 1.2.\cite{bono+09cpc}

\subsection{Pulling simulations}

\begin{figure}[t]
\includegraphics[width=0.9\columnwidth]{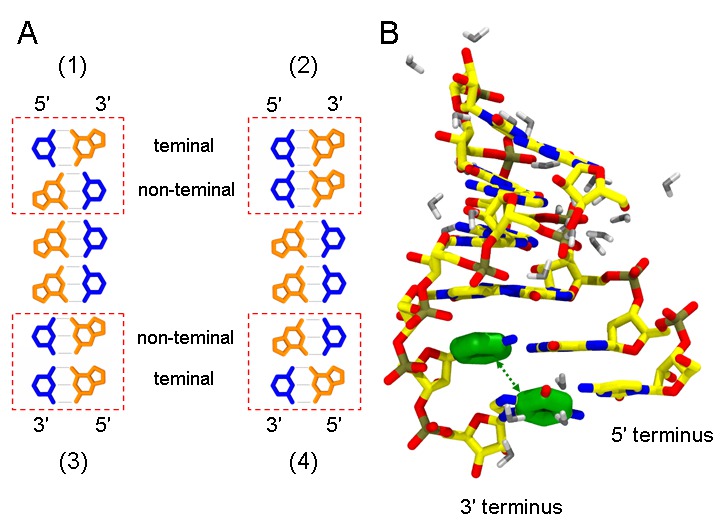}

\caption{\label{fig:double-helix}RNA double helix. A) Schematic view of the
combinations (red dotted boxes) of guanine (orange) and cytosine (blue) base-pairing
and -stacking investigated. B) Structural representation of the RNA
duplex in water; the distance between the center
of mass (green arrow) of the six-membered ring atoms (thick green
sticks) of two stacked bases was used as collective variable for the
pulling simulations.}
\end{figure}

The starting configurations for the pulling simulations were randomly
sampled from the corresponding 30 ns-long runs. The distance between
the center of mass of two stacked bases (\ref{fig:double-helix}B)
was used as pulled collective variable (CV), and thus harmonically
restrained to a constant-velocity moving point, starting at a position
equal to the equilibrium average of the CV and pulling it by 0.75
nm in 1.5 ns. This resulted in a biasing potential equal to $V_{\mbox{bias}}(q,t)=\frac{k}{2}\left[s(q(t))-\left(s_{0}+vt\right)\right]^{2}$
where $k=1200$kcal/mol/nm$^{2}$ is the spring constant of the restraint,
$q$ are the microscopic coordinates, $s(q)$ is the CV value for
those coordinates, $s_{0}$ is the initial position of the restraint,
$v$ is the pulling velocity and $t$ is the time.

\begin{figure}[!t]
 \includegraphics[width=0.9\columnwidth]{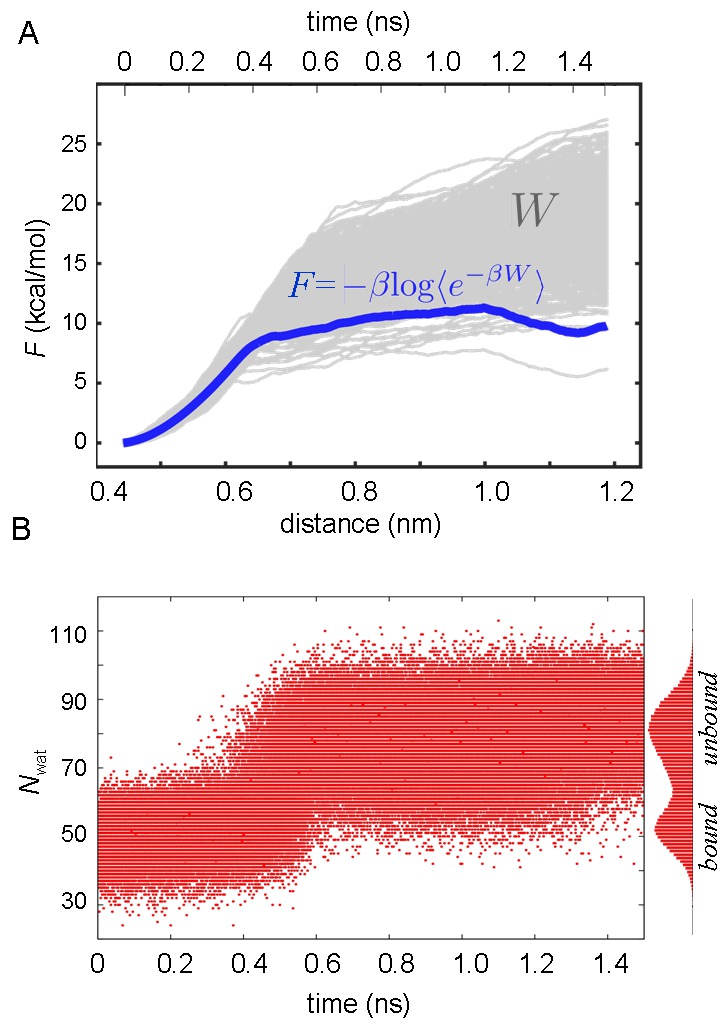}

\caption{\label{fig:work}\textcolor{black}{Typical nucleobase unbinding process,
obtained by pulling along the distance between the center of mass
of two stacked bases. A) Mechanical work (W) performed (gray plot)
and its exponential average (blue plot) as in Jarzynski equality,
plotted as a function of the restraint position. The distance is practical
for biasing the system but hardly allowed defining the bound and unbound
states. B) Number of water molecules coordinating the unbinding nucleobase ($N_{\textrm{wat}}$) }
as a function
of time (main panel), and its probability distribution (right panel).
The coordination with water is an useful metrics for identifying the
bound and unbound states.}
\end{figure}

The mechanical work done during the process was obtained by integrating
the force exerted on the system along the biased reaction coordinate.
After collecting about 400 realizations for each nucleobase-unbinding
process, the Jarzynski nonequilibrium work theorem\cite{jarz97prl}
was exploited to discount the dissipated work and to reconstruct \textcolor{black}{the
free-energy profile as a function of the restraint distance ($s_{0}+vt$).}
Although employing Jarzynski's equality in principle allows unbiased
free-energy differences to be estimated, its direct application is
limited by the number of collectable realizations as well as by the
complexity of the system.\cite{gore+2003pnas} A typical \textcolor{black}{free-energy
profile} is shown in \ref{fig:work}A as a function of the \textcolor{black}{restraint
distance}. The blue plot shows how, after a steep rise, a series of
alternating shoulders and local plateaus gradually brought the system
to higher free-energy states. Moreover, between distances ranging
from 1 to 1.2 nm (for the exemplified system), the profile was strongly
dominated by an outlier low-work realization and the difference in
reconstructing the free-energy profile with or without the outlier
was more than 3kcal/mol (see SI).

Within this framework, there was no clean way to automatically detect
when the nucleobase had reached the unbound configuration, and it
was difficult to avoid systematic errors in the comparison of many
profiles with small differences as we were interested in. To tackle
this problem, we decided to analyze our simulations in terms of CVs
different from the one used for the pulling. This \emph{a posteriori}
analysis could be done quickly, as a post-processing, and allowed
us to choose optimal CVs capable of describing in an user-independent
manner all the unbinding events.

\subsection{Reweighting scheme}

To project the free-energy landscape on putative CVs we devised a
proper reweighting scheme. Whereas suitable schemes have been proposed
to reweight other types of nonequilibrium simulations (e.g.~Ref.~\cite{bono+09jcc}),
a reweighting algorithm for steered MD has not been reported.\textcolor{black}{{}
For a different purpose, Hummer and Szabo developed a method which
enables the reconstruction of the free energy as a function of the
pulled coordinate.}\cite{humm-szab01pnas,gupt+11natphy}\textcolor{black}{{}
Here, we generalize this scheme so as to compute the free energy as
a function of any }\textcolor{black}{\emph{a posteriori}}\textcolor{black}{{}
chosen variable.}

Two different sorts of bias affect the steered MD trajectories and
needed to be removed: (a) the nonequilibrium nature of the pulling
and (b) the presence of artificial harmonic restraints on the pulled
CV. The nonequilibrium bias is removed by noticing that the equilibrium
probability $P_{\mbox{eq}}(q,t)$, for a restraint statically kept
in its position at time $t$, can be obtained from the non-equilibrium
one $P_{\text{neq}}^{(i)}(q,t)$ as observed in the $i$-th trajectory
exploiting a relation first reported by Crooks:\cite{croo00pre,will-evan10prl}

\begin{equation}
P_{\mbox{eq}}(q,t)=\sum_{i}e^{-\beta\left[W_{i}(t)-F(t)\right]}P_{\text{neq}}^{(i)}(q,t)\,,\end{equation}
where $W_{i}(t)$ is the work done on the $i$-th trajectory up to
time $t$ and $\beta=1/k_{B}T$ is the inverse thermal energy. Here
the free energy $F(t)$ represents the normalization factor corresponding
to the instantaneous position of the moving restraint at time $t$.
The bias of the harmonic restraint can then be removed by applying
the weighted-histogram analysis method.\cite{kuma+92jcc} Whereas
weighted histograms are traditionally used to combine independent
simulations performed with different static biasing potentials, here
we used it to combine snapshots obtained at different stages of the
pulling, thus writing the unbiased equilibrium probability as\begin{equation}
P_{\mbox{u}}(q)\propto\frac{\int_{0}^{\tau}dtP_{\text{eq}}(q,t)}{\int_{0}^{\tau}dte^{-\beta\left[V(q,t)-F(t)\right]}}\,,\end{equation}
where $\tau$ is the length of each pulling simulation. Finally, the
free energy as a function of an arbitrary, \emph{a posteriori} chosen
CV $\bar{s}$ is defined as $F(\bar{s})=-k_{B}T\log\int dqP_{u}(q)\delta\left(\bar{s}-\bar{s}(q)\right)$.
The scheme described so far closely resembles the one used by Hummer
and Szabo.\cite{humm-szab01pnas} \textcolor{black}{However, it is
conceptually different, as here the free energy can be reconstructed
also with respect to a variable different from the pulled one. Thus,
it potentially enables the disclosure and characterization of otherwise hidden features of the investigated
process.} To further simplify the
data manipulation and to avoid building multidimensional histograms,
with a further dependence on technical choices such as binning size,
we recast our approach assigning a weight to each of the sampled configurations,
in the same spirit as in Ref.~\cite{soua-roux01cpc}. After simple
manipulation, the weight can be shown to be equal to\begin{equation}
w_{i}(t)\propto\frac{e^{-\beta\left[W_{i}(t)-F(t)\right]}}{\int_{0}^{\tau}dt'e^{-\beta\left[V\left(q_{i}(t),t'\right)-F(t')\right]}}\,.\label{eq:weights}\end{equation}
 The normalization factor for each time, $F(t)$, is then computed
iteratively up to convergence as $e^{-\beta F(t)}=\sum_{i}\int_{0}^{\tau}dt'w_{i}(t')e^{-\beta V(q_{i}(t'),t)}$.
Usually a few tens of iterations are enough to converge.

In summary, in
our reweighting scheme we first compute the weight of each of the
configurations saved along the MD simulations from \ref{eq:weights},
then estimate free energies as a function of any, \emph{a posteriori}
chosen CV as \begin{equation}
F(\bar{s})=-k_{B}T\log\sum_{i}w_{i}(t)\delta(\bar{s}-\bar{s}(q_{i}(t))\,.\label{eq:f-of-s}\end{equation}

\section{Results}

\begin{figure}
\includegraphics[width=0.9\columnwidth]{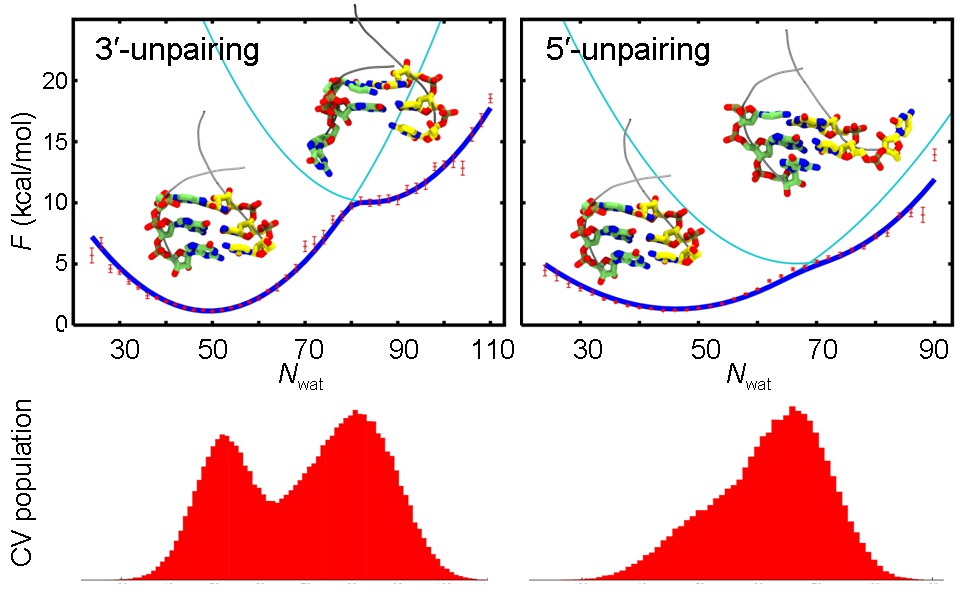}

\caption{\label{fig:fitting}Reconstruction of the free-energy profile as a
function of the number of water molecules ($N_{\textrm{wat}}$) surrounding the unbinding
base. In the left and right panels are shown typical free energy profiles
(red dots with error bar) for the \textcolor{black}{unpairing of a 3\textasciiacute{}-strand guanine
and 5\textasciiacute{}-strand cytosine}, respectively. The quadratic potentials
obtained from the double-well fitting are shown in light blue color,
whereas their combination {[}\ref{eq:combination}{]} is in blue.
Underneath each panel, the \textcolor{black}{unnormalized} population
of the CV is also shown.}
\end{figure}

Using the reweighting scheme outlined in the previous Section we were
able to investigate several order parameters. Since solvent interactions
are known to affect the conformational state of nucleic acids,\cite{cant-schi80book}
we considered the solvation of an \textcolor{black}{unbinding} base
as an effective metrics for the progression of the \textcolor{black}{underlying
process}. This choice allowed \textcolor{black}{defining} the unbinding
in a manner which was totally independent from both the \textcolor{black}{terminus}
and the specific base, and, in our explicit-solvent simulation, could
be computed as the coordination among heavy atoms of the base and
water oxygens (\ref{fig:work}B). In this metrics, the bound and unbound
states could be clearly and unambiguously identified and corresponded
to approximately harmonic basins. Sample free energies computed \textcolor{black}{as
a function of the number of water molecules coordinating the unpairing
base are shown in \ref{fig:fitting}. The free-energy
profile reconstructed along such a reaction coordinate is in no way
biased by the absolute number of coordinated water molecules which
is merely used to distinguish one configuration from the other and
to properly collect the corresponding weights along the simulation
as in \ref{eq:f-of-s}.} Then, to compute accurately the bound/unbound
free-energy differences, we fit the free-energy profiles with the combination
of two quadratic functions,\cite{humm-szab05acr,humm-szab10pnas}\begin{equation}
e^{-\beta F(\bar{s})}=\sigma_{1}^{-1}e^{-\frac{\left(\bar{s}-\bar{s}_{1}\right)^{2}}{2\sigma_{1}^{2}}-\beta F_{1}}+\sigma_{2}^{-1}e^{-\frac{\left(\bar{s}-\bar{s}_{2}\right)^{2}}{2\sigma_{2}^{2}}-\beta F_{2}}\label{eq:combination}\end{equation}
 where $F_{1}$ and $F_{2}$ are the free energies of bound and unbound
states. Both, when the two states were clearly resolvable (e.g.~\ref{fig:fitting},
left panels), and when the corresponding CV population was more overlapped,
(e.g.~\ref{fig:fitting}, right panels) the fitting procedure resulted
robust and poorly sensitive to outlier work realizations, thus enhancing
convergence of the results (e.g.~the difference in performing the
fit with or without the outlier low-work realization in \ref{fig:work}A
was less than 0.3 kcal/mol, see SI). Furthermore, this approach showed
very stable outcomes with respect to the choice of the details in
the definition of the solvation order parameter, and allowed comparing
systematically several similar situations without incurring of large
statistical errors or, worst, human biases in the interpretation of
the results.

%

Having an optimized statistical-mechanics tool able to provide
free-energy differences in a flexible and automatic manner, we pursued
a systematic step-by-step approach to investigate the feasibility
of different \textcolor{black}{opening} paths for the four possible
combinations of G-C base stacking and Watson-Crick pairing. The general
procedure, as outlined in \ref{fig:Thermodynamics-cycle},
relied on two subsequent steps: first, the Watson-Crick base pair
was partially opened by the \textcolor{black}{unpairing} of the base
on either the 5\textasciiacute{} or 3\textasciiacute{} \textcolor{black}{terminus};
second, the resulting dangling intermediate, on the 3\textasciiacute{}
or 5\textasciiacute{} \textcolor{black}{terminus} respectively, was
unstacked and the base pair opening completed.

\begin{table}
\caption{\label{tab:Context-dependent-base-unbinding}Context-dependent base-unbinding 
free energy (kcal/mol)
\textcolor{black}{
corresponding to the elementary steps shown in \ref{fig:Thermodynamics-cycle}}.
}
\begin{tabular}
{@{\extracolsep{\fill}}rrrrrr}
\multicolumn{1}{l}{Construct}& 
\multicolumn{4}{c}{Opening steps}\cr
\hline
\multicolumn1c{$n$}&
\multicolumn1c{$3b$}&\multicolumn1c{$5b$}&
\multicolumn1c{$5a$}&\multicolumn1c{$3a$}&\cr
\hline\cr
\vspace{1.5 mm}
(1)$\textrm{\ensuremath{{\textrm{5\textasciiacute-CG}..\atop \textrm{3\textasciiacute-GC..}}}}$ \quad & 7.6 & 0.9 & 2.6 & 5.6\cr
\vspace{1.5 mm}
(2)$\textrm{\ensuremath{{\textrm{5\textasciiacute-CC..}\atop \textrm{3\textasciiacute-GG..}}}}$ \quad & 8.9 & 0.3 & 3.9 & 4.5\cr
\vspace{1.5 mm}
(3)$\textrm{\ensuremath{{\textrm{5\textasciiacute-GG..}\atop \textrm{3\textasciiacute-CC..}}}}$ \quad & 7.0 & 2.2 & 4.9 & 3.9\cr
\vspace{1.5 mm}
(4)$\textrm{\ensuremath{{\textrm{5\textasciiacute-GC..}\atop \textrm{3\textasciiacute-CG..}}}}$ \quad & 7.4 & 2.0 & 5.7 & 2.8\cr
\hline
\end{tabular}
\end{table} 

%

The relative stability of putative intermediates involved in the opening
of a base pair was estimated from the individual base-unbinding free
energies (\ref{tab:Context-dependent-base-unbinding} and \ref{fig:Thermodynamics-cycle}).
For all the considered combinations,
the difference in basepair-opening free energy computed biasing the
system along path (a) and (b) in \ref{fig:Thermodynamics-cycle} was lower than 1kcal/mol.
For sake of clarity, it should be reminded that a finite number of
unidirectional pulling simulations performed within a Jarzynski-like
scheme are known to provide overestimates of absolute free-energy
differences.\cite{gore+2003pnas} However, highly accurate estimates
of unbinding constants were not needed to characterize the \textcolor{black}{strand
separation} mechanism and the free-energy differences we estimated
were exploited as a quantitative tool to assess the relative stability
of different configurations.

\section{Comparison with experiments}

Below we discuss the results of the first and second unbinding steps
(\ref{fig:Thermodynamics-cycle}), and compare them with crystal structures
conformer distributions,\cite{moha+2009jpcb,moha+2010jacs} relative
population of stacked/unstacked bases detected by femtosecond time-resolved
fluorescence spectroscopy,\cite{Liu+08bioc} and thermodynamic data
based on dsRNA melting experiments.\cite{sugi+87bioc,turn+88arbb,bloo+00book}
The consistency with experimental observations and the capability
of our simulations to complement those results are highlighted.

The more general outcome arising from the comparison of free-energy
differences is that the paired base on the 5\textasciiacute{} \textcolor{black}{terminus}
always interacted more weakly than the complementary base on the 3\textasciiacute{}
\textcolor{black}{terminus} (steps 5a, 3b in \ref{tab:Context-dependent-base-unbinding}
and \ref{fig:Thermodynamics-cycle}). This could be directly related
to the A-form helical geometry of RNA in which the bases at the 5\textasciiacute{}
end of a ss-ds junction are less buried into the neighboring environment
and expose a wider portion of their surface to water molecules, thus
facilitating fraying events. The different stability of the nucleobase
on the 5\textasciiacute{} \textcolor{black}{terminus} can be reflected
in the probability of observing a certain type of blunt closing base
pair at ss-ds junctions. In this context, the stronger interaction
was estimated for the 5\textasciiacute{}-guanine in $\textrm{\ensuremath{{\textrm{5\textasciiacute-GC..}\atop \textrm{3\textasciiacute-CG..}}}}$(construct
\textbf{4}) which was $\sim$1.7 kcal/mol weaker than the complementary
cytosine on the 3\textasciiacute{}-\textcolor{black}{terminus}. Accordingly,
the combination $\textrm{\ensuremath{{\textrm{5\textasciiacute-GC..}\atop \textrm{3\textasciiacute-CG..}}}}$(construct
\textbf{4}) is the most abundant closing base-pair pattern observed
at ss-ds junctions in large rRNA crystal structures.\cite{moha+2009jpcb}
It can be further noticed that among the dangling ends (steps 5b and
3a in \ref{tab:Context-dependent-base-unbinding} and \ref{fig:Thermodynamics-cycle})
the most stable ones are those on the 3\textasciiacute{} \textcolor{black}{terminus},
consistently with ultrafast spectroscopy experiments which have detected
a large subpopulation of stacked conformers for a 3\textasciiacute{}-dangling
fluorescent purine probe, while only a relatively small one for a 5\textasciiacute{}-dangling
purine probe.\cite{Liu+08bioc} In particular, we found the most stable
3\textasciiacute{}-dangling end in construct \textbf{1} ($\textrm{\ensuremath{{\textrm{5\textasciiacute-}{}^{\textrm{C}}\textrm{G..}\atop \textrm{3\textasciiacute-}\textrm{GC..}}}}$)\textbf{,}
which has also been counted as the most common dangling end pattern
in rRNA crystal structures.\cite{moha+2009jpcb} Further agreement
can be found considering dsRNA optical melting experiments which have
shown that single-nucleotides overhanging at 3\textasciiacute{}-ends
of \textcolor{black}{an} RNA helix increase the stability of the duplex
in a sequence-dependent manner. Notably, such a stabilization has
been interpreted as the capability of the 3\textasciiacute{} dangling
ends to stack over the hydrogen bonds of the closing base pair protecting
them from water exchange.\cite{isak-chat05bioc} Those 3\textasciiacute{}-dangling
bases which are more likely stacked would thus provide a larger \textcolor{black}{contribution}
to duplex stabilization. In this light, the trend that we observed
in the \textcolor{black}{unstacking} energies of the four dangling constructs
($\textrm{\ensuremath{{\textrm{5\textasciiacute-}{}^{\textrm{C}}\textrm{G..}\atop \textrm{3\textasciiacute-}\textrm{GC..}}}}$>$\textrm{\ensuremath{{\textrm{5\textasciiacute-}{}^{\textrm{C}}\textrm{C..}\atop \textrm{3\textasciiacute-}\textrm{GG..}}}}$>$\textrm{\ensuremath{{\textrm{5\textasciiacute-}{}^{\textrm{G}}\textrm{G..}\atop \textrm{3\textasciiacute-}\textrm{CC..}}}}$>$\textrm{\ensuremath{{\textrm{5\textasciiacute-}{}^{\textrm{G}}\textrm{C..}\atop \textrm{3\textasciiacute-}\textrm{CG..}}}}$)
is in agreement with duplex stabilization observed in dsRNA melting
experiments.\cite{sugi+87bioc,turn+88arbb,bloo+00book} \textcolor{black}{It
should be remarked that the duplex stabilization induced by 5\textasciiacute{}-dangling
ends might not reflect the stacking energy of the dangling end itself
because of its small overlap with the hydrogen bonds of the closing
base pair. Further discussion on the comparison of computed and experimental
dangling-end stabilities can be found in the SI.}

Summarizing, the unbinding of the base on the \textcolor{black}{5\textasciiacute{}-strand} was, in
all the considered cases, favored over the unbinding of the complementary
\textcolor{black}{3\textasciiacute{}-strand} base. Whereas the relative \textcolor{black}{probability}
of 3\textasciiacute{}- and 5\textasciiacute{}-unbinding event
can be modulated by the sequence, the general trend remains unchanged.

\begin{figure}
\includegraphics[width=0.9\columnwidth]{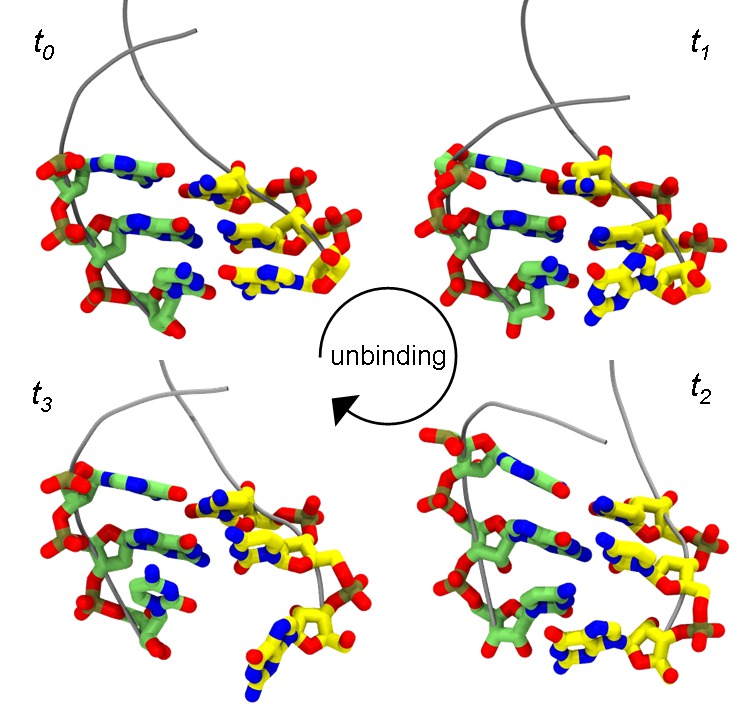}

\caption{\label{fig:Snapshots}Snapshots sampled from the opening of a base
pair. The unbinding base (here a \textcolor{black}{5\textasciiacute{}-terminal} guanine in construct
\textbf{4}) \textcolor{black}{could transiently} stack over the dangling
end of the opposite strand ($t{}_{2}$).}
\end{figure}

From a structural standpoint, the \textcolor{black}{unpairing} could
proceed through two qualitatively different paths: one in which the
twisting and breaking of Watson-Crick hydrogen bonds occurred before
the \textcolor{black}{rupture of stacking interactions}; the other,
in which the unbinding followed the concerted rupture of both hydrogen
bonds and stacking interactions with, in some cases, the unbinding
base stacking over the dangling end of the opposite strand (\ref{fig:Snapshots}).
Similar unbinding geometries have also been described in other studies.\cite{poho+90ijsa,haga+03pnas}
In our simulations, these intermediate states occurred with a context-dependent
frequency. For instance, along the unbinding pathway the \textcolor{black}{5\textasciiacute{}-terminal} guanine
in $\textrm{\ensuremath{{\textrm{5\textasciiacute-GC..}\atop \textrm{3\textasciiacute-CG..}}}}$
(construct \textbf{4}) had $\sim$15\% of probability to stack upon
the 3\textasciiacute{}-dangling cytosine of the opposite strand.
Interestingly, this probability dropped to $\sim$3\% in $\textrm{\ensuremath{{\textrm{5\textasciiacute-GG..}\atop \textrm{3\textasciiacute-CC..}}}(construct \textbf{3})}$.
Such an inter-strand stacking pattern (\ref{fig:Snapshots}, panel
$t_{2}$), exchanging with the conventional stacking of a paired base,
could account for the delayed quenching of fluorescence detected by
ultrafast fluorescence spectroscopy for a construct similar to $\textrm{\ensuremath{{\textrm{5\textasciiacute-GC..}\atop \textrm{3\textasciiacute-CG..}}}}$
(construct \textbf{4}).\cite{Liu+08bioc}

As final experimental evidences which corroborate our results, Xia
and co-workers\cite{Liu+08bioc} have reported that the dynamic behavior
of a 3\textasciiacute{}-terminal purine is not affected by the presence
of the opposite complementary base. Viceversa, the conformational
dynamics of a 5\textasciiacute{}-terminal purine is drastically
influenced by the presence of an opposite 3\textasciiacute{}-\textcolor{black}{terminal} pyrimidine
which would be likely stacked and potentially able to shift the population
of the \textcolor{black}{complementary 5\textasciiacute{}-terminal} base towards a paired and stacked
ensemble. Consistently with our systematic study, these data depict
the formation of a stable base pair as generally driven by the stacking
of the \textcolor{black}{3\textasciiacute{}-terminal base}, and then
by the energy gained by the system from both the stacking of the \textcolor{black}{5\textasciiacute{}-terminal
base} and Watson-Crick hydrogen-bonds.

\section{Biological implications}

\begin{figure*}
\includegraphics[width=1\textwidth]{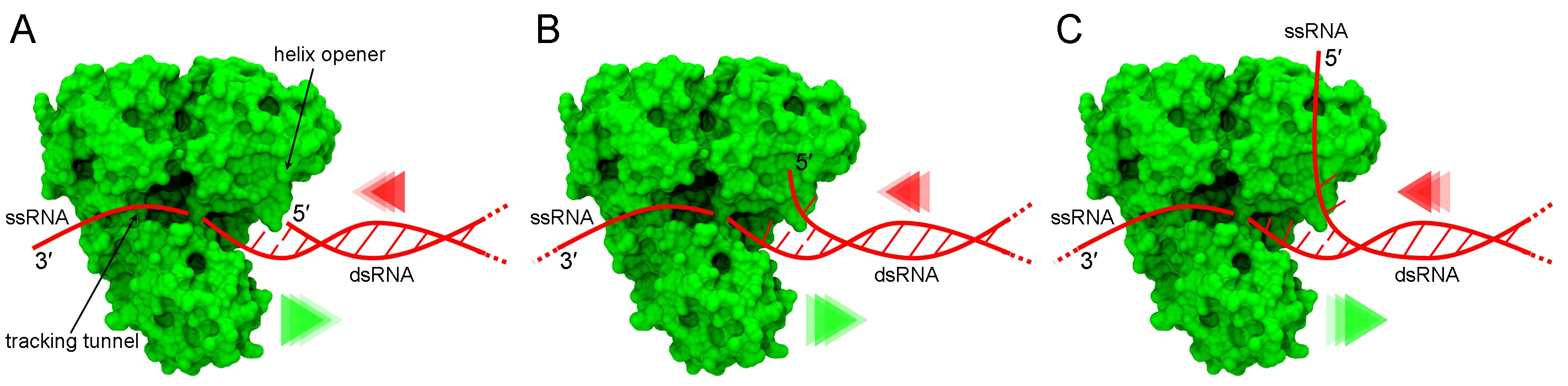}
\textcolor{black}{
\caption{
\label{fig:Helicase}
Model for RNA unwinding catalyzed by NS3 helicase.
A) The single-stranded 3\textasciiacute{} terminus
is loaded into the tracking tunnel.
B) The 5\textasciiacute{} terminus is mechanically displaced
by the helix opener.
C) The helicase proceeds
with 3\textasciiacute{} to 5\textasciiacute{} directionality,
displacing the 5\textasciiacute{}-strand.
}
}
\end{figure*}
The motion picture of
\textcolor{black}{duplex separation}
emerging from the outcome of our simulations complements and augments
with dynamic details and energetic considerations the helix propagation
model based on the analysis of static 3D structures.\cite{moha+2009jpcb}
Our computations link experimental data from different fields creating
a common reading frame among them. Taken together, these results suggest
that RNA unwinding occurs by a stepwise process in which the probability
of unbinding of the base on the 5\textasciiacute{} \textcolor{black}{strand}
is significantly higher than that on the 3\textasciiacute{} \textcolor{black}{strand}.
What could be the biological implications of this finding?

When considering the RNA as the substrate of molecular motors such
as helicases and other remodeling enzymes, the results could likely
be interpreted from an evolutionary point of view which could allow
deciphering the basis of the evolutionary pressure responsible for
the unwinding mechanism catalyzed by RNA-duplex processing enzymes.

The RNA unwinding catalyzed by helicases is coupled to adenosine triphosphate
(ATP) binding and hydrolysis. The underlying mechanism would reasonably
minimize the use of ATP, especially in a low-nutrient environment.
Provided that the intrinsic RNA dynamics implies that at ss-ds junctions
the unbinding of the 5\textasciiacute{}-strand
base is favored over the unbinding of the complementary 3\textasciiacute{}-strand base,
an enzymatic unwinding model would include a mechanism in which the separation
of the two complementary strands is accomplished by acting on the
weakest portion, i.e.~the 5\textasciiacute{}-strand
base. Thus, an ancestral enzyme using the 3\textasciiacute{}-strand
as running track rail (with 3\textasciiacute{} to 5\textasciiacute{}-directionality)
without perturbing its conformation and causing the displacement of
the 5\textasciiacute{}-strand by mechanical exclusion
could satisfy some energy-saving requirements.

The viral RNA helicase NS3 of hepatitis C virus, which is a prototypical
DEx(H/D) RNA helicases essential for viral replication, could satisfy
the above mentioned requirements.\cite{jank-fair07cosb,pyle08arb}
NS3 is a potentially relevant drug target and has been structurally
and functionally characterized in various contexts.\cite{dumo+06nat,buet+07nsmb,gu-rice10pnas,rane+10jbc}
It unwinds duplexes by first loading onto a \textcolor{black}{single-stranded 3\textasciiacute{}-terminus}
region and then processively translocating with 3\textasciiacute{}
to 5\textasciiacute{} directionality along this loading strand,
thereby peeling off the complementary 5\textasciiacute{}-strand
bases. In particular, the 3\textasciiacute{}-strand
would migrate through a tracking ssRNA tunnel running within the protein
whereas the complementary 5\textasciiacute{}-strand
is forced towards the back of the protein by the {}``helix opener''
hairpin (\ref{fig:Helicase}).\cite{Luo+08emboj,sere+09jbc} In light of the free-energy
calculations discussed above, it could be suggested that this mechanism
has been optimized according to the intrinsic RNA unwinding dynamics
disclosed in this work.

Arguably, the processing machineries are being constantly shaped by
the evolutionary pressure of a plethora of (often unknown) factors
contributing to the optimization of metabolism in the whole living
system, rather than to the local biochemical process. As a consequence,
the preference for a well-defined RNA processing directionality cannot
be ubiquitously observed.\cite{jank-fair07cosb,pyle08arb}

We speculate that \emph{all} the biochemical processes involving RNA
in which directionality plays a role (e.g.~transcription), could be
related to the energetics of RNA double helix forming and fraying
discussed in this Article.

\section{Conclusions}

This study lays down the basis for the molecular-level understanding
of intrinsic RNA dynamics and its role in function. The asymmetric
behavior of the 3\textasciiacute{}- and 5\textasciiacute{}-strand
could be responsible for the directionality observed in RNA processing.
From a computational perspective, the approach we introduced can be
generalized to analyze any kind of (un)binding event. Indeed, it allowed
the free-energy landscape to be reconstructed along different reaction
coordinates and the unbinding energies to be easily computed in an
automatic and user-independent manner, therefore removing statistical
and human biases. We foresee the application of our approach to a
wider range of molecular systems, including the typical ligand-target
complex faced in drug discovery.\cite{jorg10nat}


\section{Acknowledgment} 

We thank Francesco Di Palma, Rolando Hong and Vittorio Limongelli for critically reading
the manuscript and an anonymous referee for several useful suggestions. We acknowledge the CINECA Award N.~HP10BLIT9Z, 2011
for the availability of high performance computing resources and MIUR
grant ``FIRB - Futuro in Ricerca'' N.~RBFR102PY5 for funding. Developed
software is available on request. 

\begin{suppinfo}

Details of methodology and computations. Unbinding free-energy profiles
for both terminal and non-terminal base pairs. \textcolor{black}{Further
discussion on the comparison of computed and experimental dangling-end
stabilities.}

\end{suppinfo}

\providecommand*\mcitethebibliography{\thebibliography}
\csname @ifundefined\endcsname{endmcitethebibliography}
  {\let\endmcitethebibliography\endthebibliography}{}



\end{document}